\documentclass[pre,amsmath,amssymb,preprint,superscriptaddress]{revtex4}

\newcommand{\apjl}{\mbox{\it Astrophys. J.}}
\newcommand{\mnras}{\mbox{\it Mon. Not. R. Astron. Soc.}}
\newcommand{\pasp}{\mbox{\it Publ. Astr. Soc. Pacific}}
\newcommand{\physrep}{\mbox{\it Phys. Rep.}}
\newcommand{\ssr}{\mbox{\it Space Science Rev.}}
\newcommand{\beq}{\begin{equation}}
\newcommand{\eeq}{\end{equation}}
\newcommand{\ba}{\begin{array}}
\newcommand{\ea}{\end{array}}

\usepackage{amsmath}    
\usepackage{amsfonts}
\usepackage{amssymb}
\usepackage{graphicx}   
\usepackage{dcolumn}
\usepackage{bm}
\usepackage{color}

\begin{document}

\title{On the Formation and Properties of Fluid Shocks and Collisionless Shock Waves in Astrophysical Plasmas}

\author{A. Bret}
\affiliation{ETSI Industriales, Universidad de Castilla-La Mancha, 13071 Ciudad Real, Spain}
 \affiliation{Instituto de Investigaciones Energ\'{e}ticas y Aplicaciones Industriales, Campus Universitario de Ciudad Real,  13071 Ciudad Real, Spain.}

\author{A. Pe'er}
\affiliation{Department of Physics, University College Cork, Cork, Ireland}

\date{\today }

\begin{abstract}
When two plasmas collide, their interaction can be mediated by
collisionless plasma instabilities or binary collisions between
particles of each shell. By comparing the maximum growth rate of the
collisionless instabilities with the collision frequency between
particles of the shells, we determine the critical density separating
the collisionless formation from the collisional formation of the
resulting shock waves. This critical density is also the density
beyond which the shock downstream is field free, as plasma
instabilities do not have time to develop electromagnetic patterns. We
further determine the conditions on the shells’ initial density and
velocity for the downstream to be collisional. If these quantities
fulfill the determined conditions, the collisionality of the
downstream also prevents the shock from accelerating particles or
generating strong magnetic fields. We compare the speed of sound with
the relative speed of collision between the two shells, thus
determining the portion of the parameters space where strong shock formation
is possible for both classical and degenerate plasmas. Finally, we discuss the
observational consequences in several astrophysical settings.
\end{abstract}

\maketitle

\section{Introduction}

Shock waves are among the most ubiquitous and most studied physical
phenomena. They exist in  many different astronomical objects, on
very many different scales. They play a major role in shaping the
observed signal of various objects, providing (1) a natural way of
depositing kinetic energy; (2) the necessary conditions for
acceleration of particles to high energies, non-thermal distribution;
and furthermore, (3) shock waves may be responsible for generating
strong magnetic fields \cite{Bell78, Silva+03, Spit08, SS11}.

Shock waves may come in two flavors. In a neutral fluid, kinetic
energy dissipation at the shock front is provided by binary
collisions \footnote{We focus the discussion here on shocks that
  propagate in an environment in which the energy density of
  radiation downstream can be neglected with respect to the particle
  thermal energy.}. As a result, the shock front is a few
mean-free-paths thick \cite{Zeldovich}, and the shock is
``collisional''.  In charged plasma, on the other hand, instabilities
prompted by collective behavior can equally mediate shock waves and
provide the kinetic energy dissipation at the front
\cite{Petschek1958,Sagdeev66}. In this case, the shock front can be
several orders of magnitudes shorter than the mean-free-path for
binary collisions \cite{BambaApJ2003}. These shocks have been dubbed
``collisionless shocks''.

While there were still doubts about the very existence of
collisionless shocks in the late 1980's \cite{sagdeev1991}, in-situ
observations of the earth bow-shock, for example, have definitely cast
them out \cite{PRLBow1,PRLBow2}. Because they are collisionless, these
 shocks are formed through collective plasma instabilities,
on the time scale of these instabilities
\cite{BretPoP2013,BretPoP2014}.

The absence of close collisions allows particles to gain energy
without sharing it with others. As a result, collisionless shocks
are excellent particle accelerators \cite{Blandford1987,Marco2016}, as
opposed to collisional shocks \cite[e.g.,][]{Longair11}.

In view of the properties which derive from the absence of collisions,
one can wonder about the conditions required for collisionless-ness to
be fulfilled. The goal of this paper is to investigate 1/ the
conditions required for the shock formation to be collisionless, and
2/ the conditions required for the downstream region to be so.

Regarding the first item, the nature of the shock formation
determines the time scale on which it forms. In the collisionless
regime, the two colliding shells start passing through each other. The
overlapping region quickly turns unstable, generating a turbulence
which blocks the flow and triggers the shock formation
\cite{BretPoP2013,BretPoP2014}. Binary collisions between the
particles of each shell can be neglected if the (average) collision
frequency $\nu_{ss}$ is much smaller than the growth-rate $\delta$ of
the fastest counter-streaming instability involved in the overlapping
region. On the other hand, if $\delta \ll \nu_{ss}$, binary collisions
govern the dynamics of the shells encounter. We thus find that a
quantitative investigation of the interface between the collisional
and the collisionless regimes, comes down to comparing $\delta$ and
$\nu_{ss}$. Note that such an endeavor only makes sense in a plasma,
for in a non-ionized collisionless medium, counter-streaming flows are
stable ($\delta=0$) and can only be disrupted by binary collisions.

Still for the first item, the nature of the shock formation determines
the electromagnetic patterns that will be found in the downstream,
once the shock is formed. Such patterns, like Weibel filaments
\cite{Milos2006,Huntington2015}, are the fruit of plasma instabilities
in the collisionless case. If the shock is formed through close binary
collisions, these instabilities will not grow and will not be able to
seed electromagnetic patterns in the downstream. In turn, the absence
of fields in the downstream means the absence of scattering agents for
the particles, inhibiting their acceleration.

As for the ability of the shock to accelerate particles, we point out that the time scale required for Fermi acceleration is much longer than that required for shock formation. Therefore, even a weak collisionality could allow for collisionless shock formation, while suppressing acceleration. As a result, the limit for acceleration set below, namely by ``inter-shell collision frequency = growth rate'', could indeed be an upper bound, as acceleration could be cancelled even slightly before this threshold.

Regarding the second shock item investigated if this paper, namely,
the collisionality of the downstream, it also determines whether or
not the shock, once formed, is capable of accelerating particles. The
reason for this comes from the fact that particle acceleration in a
Fermi process results from back-and-forth motions around the shock
front \cite{Spitkovsky2008a}. This is only possible if both the
upstream and the downstream are collisionless, so that particles can
nearly freely travel between each region, without exchanging energy
with the others. But if the downstream happens to be collisional,
particles will be trapped inside as soon as they enter it. They will
remain embedded into the downstream flow, constantly exchanging energy
with the others, and unable to keep it or to close any Fermi
acceleration cycle.

While the shock compresses the gas, it also compresses any parallel magnetic field. Thus, while the density in the downstream region is higher than in the upstream region, particles accelerated by a Fermi mechanism can spend more time in the upstream region. Thus, in determining the ability of the shock to support acceleration of particles to high energies, one needs to probe the conditions in both the upstream and downstream regions.

Nonetheless, as the compression ratio of the density is similar to that of the parallel component of the magnetic field, the ratio of Larmor radii in the upstream and downstream regions cannot exceed the ratio of downstream to upstream densities. As a consequence, the analysis of the conditions at the upstream region is not expected to affect the results obtained by analysing the downstream region alone by a factor larger than two.

This paper is structured as follow. We begin by considering pair
plasmas in Section \ref{sec:2}. In Section \ref{sec:collfreq} we
calculate the collision frequency $\nu_{ss}$ for close Coulomb
collisions between particles of the two pair shells. We then compute
in Section \ref{sec:insta} the growth-rate $\delta$ of the fastest
growing collisionless mode. In Section \ref{sec:compa}, $\nu_{ss}$ and
$\delta$ are compared, allowing us to determine the portions of the
phase space $(\gamma_0,n_0)$ (initial Lorentz factor and density of
the colliding plasma shells) where the shock formation is mediated by
collisionless plasmas effects, or inter-shells binary
collisions. Finally, Section \ref{sec:down_coll} derives the
requirements on $(\gamma_0,n_0)$ for the downstream of the shock, once
formed, to be collisional.

We then turn to electron/proton plasmas in section
\ref{sec:case_ep}. We conduct similar calculations as for the pair
plasmas case, emphasizing the qualitative difference that results from
the difference in instability growth rate in this scenario. Following
these theoretical derivations, we explore the limits of the small
velocity spread approximation within each shell used throughout this
work in section \ref{sec:T=0}. By equating the speed of sound in the
different regimes (classical and quantal gas, Newtonian and highly
relativistic) to the speed of collision between the shells, we
constrain the parameter space region in which strong shocks can possibly
form. Finally, we discuss in section \ref{sec:obs} the conditions that
exist inside several astronomical objects, and the applicability of
the theory to these various objects, before we reach our conclusions
in section \ref{sec:conclusions}.

\section{Colliding pair plasma shells}
\label{sec:2}

\begin{figure}
  \begin{center}
   \includegraphics[width=.7\textwidth]{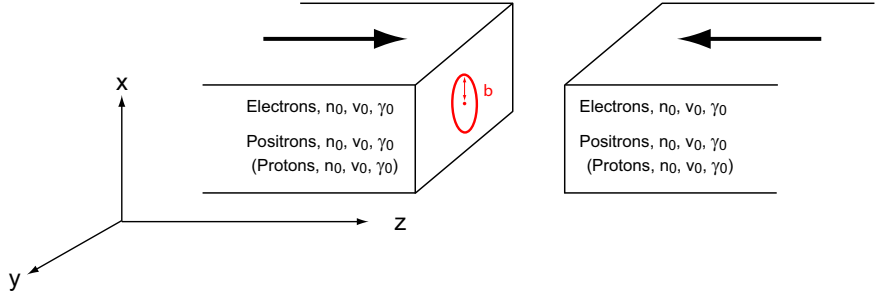}
  \end{center}
\caption{System considered: two counter-streaming plasmas
  collide. They are initially cold and symmetric, with electronic
  density $n_0$ in the laboratory frame. The first part of the article deals with pair
  plasmas, while the second part deals with electron/proton
  plasmas.}
\label{fig:setup}
\end{figure}

We begin by considering the scenario of two symmetric plasma shells
composed of electron-position pairs heading toward each other, as
pictured in Figure \ref{fig:setup}. Each shell is initially cold (see discussion in Section \ref{sec:T=0}), with
initial (lab frame) density $n_0$ and Lorentz factor
$\gamma_0=(1-\beta_0^2)^{-1/2}$, where $\beta_0=v_0/c$ is the
normalized flow velocity.


\subsection{Inter-shell collision frequency}
\label{sec:collfreq}

The impact parameter for close binary Coulomb collisions is defined
via (see Equation (13.4) in \cite{jackson1998}, with deviation
$\theta=\pi/2$)
\begin{equation}\label{eq:bC}
  b_C = \frac{q^2}{\gamma_r m_e v_r^2},
\end{equation}
where $\gamma_r$ is the relative Lorentz factor of the two shells,
namely $\gamma_r = 2\gamma_0^2-1$.
%
The relative velocity of the two shells is
\begin{equation}
\label{eq:vrell}
  v_r = \frac{2\beta_0}{1+\beta_0^2}c,
\end{equation}
where clearly $\gamma_r=(1-v_r^2/c^2)^{-1/2}$. When $b_C$ becomes too
small, it has to be replaced by the relevant de Broglie length (see
Equation (5.10) in \cite{rybicki}),
\begin{equation}\label{eq:bQ}
  b_Q = \frac{\hbar}{p} = \frac{\hbar}{\gamma_r m_e v_r}.
\end{equation}
The frequency for close collisions between particles of two different shells then reads,
\begin{equation}\label{eq:coll_freq}
  \nu_{ss} = n_0 v_r \pi b^2 = n_0 \frac{2\beta_0}{1+\beta_0^2}c ~
  \pi \max (b_C, b_Q)^2.
\end{equation}

Since $b_C$ is proportional to $v_r^{-2}$ and $b_Q \propto v_r^{-1}$,
the impact parameter is classical at low velocity, and quantum at high
velocity. The two values of the impact parameters become equal at
%
\begin{equation}
  \frac{q^2}{\gamma_r m_e v_{r,eq.}^2} = \frac{\hbar}{\gamma_r m_e
  v_{r,eq.}} \Rightarrow \frac{v_{r,eq.}}{c} \equiv \beta_{r,eq.} =
\frac{q^2}{\hbar c} = \alpha,
\end{equation}
namely at sub-relativistic velocities. Here $\alpha \sim 1/137$ is the
fine structure constant.

In terms of $\beta_0$ and $\gamma_0$, equality is achieved for
\begin{eqnarray}\label{eq:v0crit}
 \gamma_{0,eq.} & \equiv & \gamma_0^* = \sqrt{\frac{\gamma_{r,eq.}+1}{2}} \sim 1.0000067, \nonumber \\
 \beta_{0,eq.}  & \equiv & \beta_0^* = \sqrt{1-1/{\gamma_0^*}^2}   \sim 0.00365.
\end{eqnarray}
%
We thus find that for $\beta_0 < \beta_0^*$, the frequency of close
inter-shells collisions is given by Equation (\ref{eq:coll_freq}) with
$\max (b_C, b_Q) = b_C$. For $\beta_0 > \beta_0^*$, it is given by
Equation (\ref{eq:coll_freq}) with $\max (b_C, b_Q) = b_Q$.


\begin{figure}
\begin{center}
 \includegraphics[width=.7\textwidth]{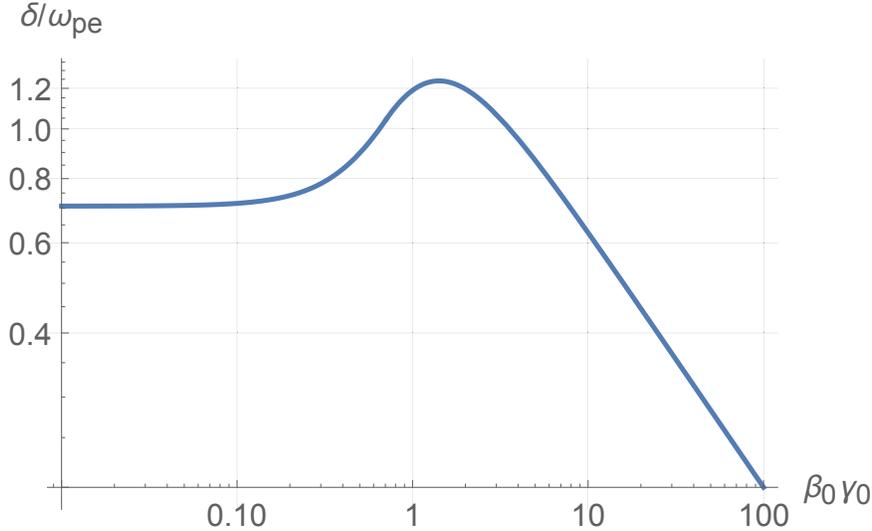}
\end{center}
\caption{Growth-rate of the fastest growing mode in terms of
  $\beta_0\gamma_0$ for cold pair plasmas
  interaction.}\label{fig:delta}
\end{figure}

\subsection{Maximum instability growth-rate $\delta$}
\label{sec:insta}

In collisionless plasmas, shock formation is triggered by the
counter-streaming instabilities that arise when the shells start
overlapping. These instabilities are numerous and can be found for
wave vectors aligned, normal, or even oblique to the flow
\cite{Watson,BretPoPReview}. For the present case, the growth-rate of
the fastest growing mode is only a function of the Lorentz factor. It
has already been determined for any $\gamma_0$ in Reference \cite{BretPoPReview};
for completeness, we here recall the main results. Noteworthily,
the forthcoming growth-rates are analytically exact.

For $\gamma_0 > \sqrt{3/2}$ ($\beta_0 > 1/\sqrt{3}$), the fastest
growing mode is the Weibel mode \cite{BretPoP2013}, with $\mathbf{k}
\perp \mathbf{v}_0$. Its growth-rate is
\begin{equation}\label{eq:gr_relat}
  \delta = 2 \frac{\beta_0}{\sqrt{\gamma_0}} \omega_{pe},
\end{equation}
where $ \omega_{pe} = \sqrt{4\pi n_0 q^2/m_e}$ is the (electron) plasma frequency.
In the opposite regime $\gamma_0 < \sqrt{3/2}$, the fastest growing
mode is oblique, with growth-rate
\begin{equation}\label{eq:gr_nr}
  \delta = (1+\beta_0^2) \sqrt{\frac{\gamma_0}{2}} \omega_{pe}.
\end{equation}
Expressions (\ref{eq:gr_relat}) and (\ref{eq:gr_nr}) are plotted
together as functions of $\beta_0\gamma_0$ in Figure
\ref{fig:delta}. The threshold $\gamma_0 = \sqrt{3/2}$ corresponds to
$\beta_0\gamma_0=1/\sqrt{2}$. The growth rate $\delta$ obtains a
maximum value $\delta/\omega_{pe}=2^{3/2}/3^{3/4} = 1.24$ for
$\gamma_0=\sqrt{3}$ ($\beta_0\gamma_0=\sqrt{2}$).

The existence of a maximum value for the growth-rate $\delta$ can intuitively be
understood as follows.  
For plasmas having Lorentz
factor $\gamma_0=\sqrt{3}$ the dominant instability is the Weibel
instability (since $\sqrt{3} > \sqrt{3/2}$).
%
This instability is driven by the repulsion of opposite currents
\cite{MedvedevApJ2005}. It relies therefore on the Lorentz force being
$\propto v_0$. As a result, it weakens at low velocities. Furthermore,
this instability weakens at high velocities as well, since $v_0$
cannot surpass $c$ while the relativistic inertia keeps increasing
with $\gamma_0$. These two features are reflected in the scaling of
the growth-rate (\ref{eq:gr_relat}), which varies like
$\beta_0/\sqrt{\gamma_0}$. With a null limit both at high and low
velocities, an intermediate extremum is necessary. Solving
$\partial\delta/\partial\gamma_0=0$ gives $\gamma_0=\sqrt{3}$.

\subsection{Comparison of collision rate $\nu_{ss}$ and instability
  growth rate $\delta$}
\label{sec:compa}

\begin{figure}
  \begin{center}
   \includegraphics[width=.7\textwidth]{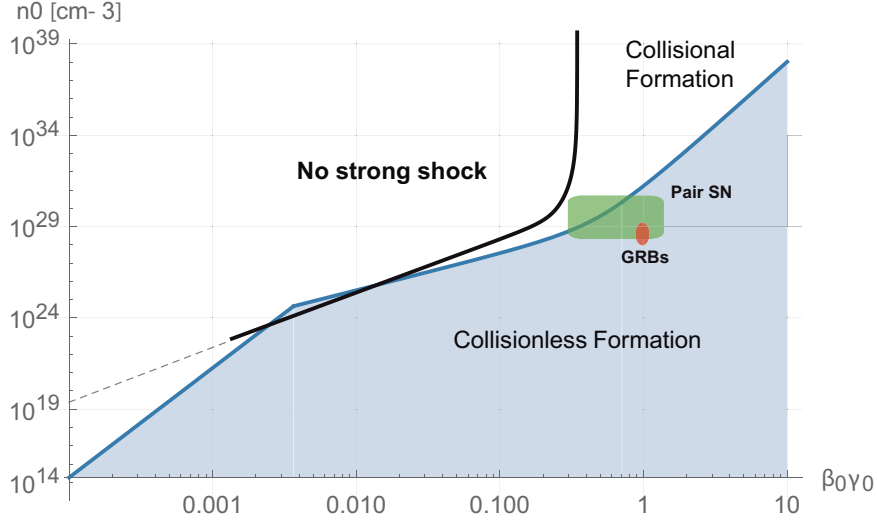}
  \end{center}
\caption{Critical density beyond which the interaction is collisional
  for pair plasmas. The blue shaded area of the phase space pertains
  to collisionless interactions. The colored regions pertain to
  astrophysical scenarios discussed in Section \ref{sec:obs}. The bold
  black line pictures the no strong shock condition discussed in section
  \ref{sec:T=0}.  As discussed in that section, for $T<T_F$ (the Fermi
  temperature) strong shocks cannot be formed above this line, for the speed
  of sound is larger than the collision speed. }
\label{fig:ncrit}
\end{figure}

In order to determine whether the shock formed is collisional or
collisionless, one needs to compare the frequency for close
Coulomb collisions with the fastest growth-rate given above. One singles
out three intervals:
\begin{enumerate}
\item
For $\beta_0<\beta_0^*$, the impact parameter is classical. The
relevant growth-rate in this regime is given by Equation (\ref{eq:gr_nr}). In this scenario, one compares Equation (\ref{eq:coll_freq}) with $b_C >
b_Q$ and Equation (\ref{eq:gr_nr}). The calculation gives
\begin{equation}
\label{eq:ratio1}
\frac{\nu_{ss}}{\delta} = \sqrt{\frac{\pi}{128}}
\sqrt{\frac{n_0}{N^*}}\frac{1}{\beta_0^3\gamma_0^{9/2}} \simeq
\sqrt{\frac{\pi}{128}} ~ \sqrt{\frac{n_0}{N^*}} ~ \frac{1}{\beta_0^3}.
\end{equation}
Here,
\begin{equation}\label{eq:N*}
N^* \equiv \left( \frac{m c^2}{q^2} \right)^3 = 4.5 \times
10^{37}~\mathrm{cm}^{-3}.
\end{equation}

\item
For $\beta_0 > \beta_0^*$ and $\gamma_0 < \sqrt{3/2}$, the impact
parameter is quantum, while the relevant growth-rate is still given by
Equation (\ref{eq:gr_nr}). Comparing Equation (\ref{eq:coll_freq}) but now with $b_C < b_Q$ and
Equation (\ref{eq:gr_nr}) gives
\begin{equation}\label{eq:ratio2}
\frac{\nu_{ss}}{\delta} = \sqrt{\frac{\pi}{8}}
\sqrt{\frac{n_0}{N_1^*}} \frac{1}{\gamma_0^{9/2}\beta_0
  (1+\beta_0^2)^2} \simeq \sqrt{\frac{\pi}{8}} ~
\sqrt{\frac{n_0}{N_1^*}} ~ \frac{1}{\beta_0}.
\end{equation}
Here,
\begin{equation}\label{eq:N1*}
N_1^* = \frac{m c^2}{q^2} a_0^{-2} = 1.26 \times
10^{29}~\mathrm{cm}^{-3},
\end{equation}
and $a_0=\hbar^2/mq^2$ is the Bohr radius.

\item
For $\gamma_0 > \sqrt{3/2}$, the impact parameter is quantum, while the
relevant growth-rate in this case is given by Equation (\ref{eq:gr_relat}). Comparing Equation (\ref{eq:coll_freq}) with $b_C < b_Q$ and Equation
\ref{eq:gr_relat} gives
\begin{equation}
\label{eq:ratio3}
\frac{\nu_{ss}}{\delta} = \frac{\sqrt{\pi}}{8}
\sqrt{\frac{n_0}{N_1^*}} \frac{1}{\gamma_0^{7/2}\beta_0^2
  (1+\beta_0^2)} \simeq \frac{\sqrt{\pi}}{16} ~
\sqrt{\frac{n_0}{N_1^*}} ~ \frac{1}{\gamma_0^{7/2}}.
\end{equation}
\end{enumerate}
One thus concludes that in all 3 cases, the ratio $\nu_{ss}/\delta$
has the form
\begin{equation}
\frac{\nu_{ss}}{\delta} = F \times \sqrt{\frac{n_0}{N^*_i}},
\end{equation}
where $N^*_i=N^*$ or $N^*_1$, depending on the scenario
considered. The value of the function $F = F(\beta_0)$ is defined
through Equations (\ref{eq:ratio1}, \ref{eq:ratio2}, \ref{eq:ratio3}), and
depends on the scenario considered. We thus conclude that the shock is collisional if
\begin{equation}\label{eq:F}
\frac{\nu_{ss}}{\delta} > 1 \Rightarrow n_0 > \frac{N^*_i}{F^2}.
\end{equation}
Figure \ref{fig:ncrit} pictures the critical density beyond which the
interaction is collisional. The lower part of the $(\gamma_0,n_0)$
phase space pertains to collisionless interactions.

\begin{figure}
  \begin{center}
   \includegraphics[width=.9\textwidth]{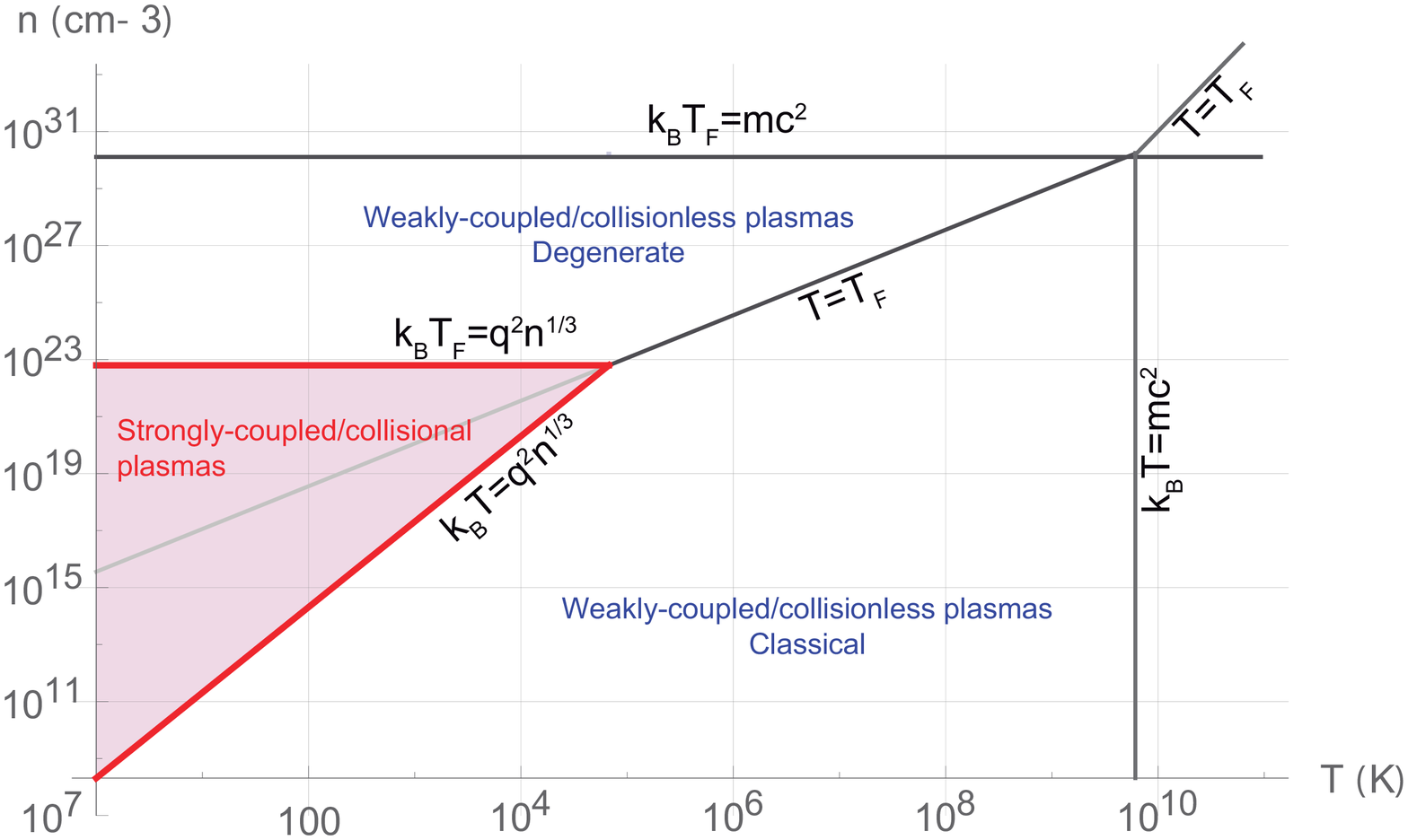}
  \end{center}
\caption{Collisionless region in the $(T,n)$ phase space. The red
  shaded area pertains to strongly coupled plasmas, i.e., collisional
  plasmas. A classical plasma becomes relativistic for $k_BT > mc^2$,
  and a degenerate plasma becomes relativistic for $k_BT_F >
  mc^2$.}
\label{fig:classi}
\end{figure}

\subsection{Conditions on $n_0$ and $\gamma_0$ for a collisionless downstream}
\label{sec:down_coll}

After the two shells collide, a shock is formed with a downstream
having density $n$ and temperature $T$. Depending on $n$ and $T$, the
downstream region can be collisionless or collisional - even if
initially the shock is formed collisionlessly.

A collisionless plasma is weakly coupled. Indeed, by definition, in a
weakly coupled plasma the kinetic energy of the plasma's particles is
much larger than the potential energy associated with Coulomb
collisions.  This, in turn, implies that close Coulomb collisions are
rare. The parallel extends even to the degenerate regime, as, for
example, the Fermi energy of a weakly coupled degenerate plasma can be
found assuming free wave functions (plane waves $\propto
e^{i\mathbf{k}\cdot\mathbf{r}}$) for the particles
\cite{ashcroft1976}.

Figure \ref{fig:classi} shows the different plasma regimes in the
downstream region, in terms of its temperature and density $(T,n)$. The line $T=T_F$ pictures
plasmas with temperature equal to the Fermi temperature. Below this
line, $T > T_F$ and the plasma is classical. It is therefore weakly
coupled if its kinetic energy is greater than the Coulomb potential,
i.e., $k_BT > q^2 n^{1/3}$ \footnote{Note that in calculating the
  Coulomb potential there is a pre-factor of order unity, which is
  omitted for clarification.}. Above this line, $T<T_F$ and the plasma
is degenerate. It is therefore weakly coupled if the Fermi energy is
greater than the Coulomb potential, $k_BT_F > q^2 n^{1/3}$
\cite{Ichimaru1982}. As the Fermi temperature $T_F$ only depends on the
density, this latter condition defines a critical density threshold,
\begin{eqnarray}\label{eq:EF}
k_BT_F &=& (3\pi^2n)^{2/3}\frac{\hbar^2}{2m_e} = q^2 n^{1/3}
\\ \Rightarrow n &>& \frac{8}{9\pi^4} \left( \frac{m_e q^2}{\hbar^2}
\right)^3 = 6.3 \times 10^{22} ~ \mathrm{cm}^{-3} \nonumber,
\end{eqnarray}
instead of an oblique line. As a result, plasmas located inside the
red triangle pictured in Figure \ref{fig:classi} are strongly coupled,
i.e., collisional. If the downstream lies in this domain, the shock
cannot accelerate particles. Note that at higher densities, the plasma
remains collisionless, somewhat counter-intuitively, as it is kept
degenerate as the densities increase, until reaching the relativistic
limit. Thus, the rest of the phase parameter space is weakly coupled,
whether for classical or quantum reasons.

One can notice that the collisional regime has an upper bound, both in
temperature and density. For $T > 10^5$ K, or $n >6.3 \times 10^{22} ~
\mathrm{cm}^{-3}$, the plasma cannot be collisional. Rather, it must be
collisionless.

Given that both $n$ and $T$ are functions of the initial density and
Lorentz factors $n_0$ and $\gamma_0$, one can determine the
requirements on the initial plasma parameters $(n_0, \gamma_0)$ that
result in collisional downstream region. From the discussion above, it
follows that there are two requirements: (i) for classical plasma
regime, it is $k_BT \geq q^2 n^{1/3}$; (ii) for the quantum border of
the collisional regime, one requires that the plasma density is $n>6.3
\times 10^{22}$~cm$^{-3}$. We therefore explore the conditions on
$n(n_0,\gamma_0)$ and $T(n_0,\gamma_0)$ that fulfill these
requirements.


\subsubsection{Conditions for the downstream to be collisional}

We derive in Appendix \ref{sec:nT} the dependence of the downstream
density and temperature $(n,T)$ on the initial density and the Lorentz
factor, $(n_0,\gamma_0)$ in the non-relativistic regime, which is the
relevant regime here.  The relations derived in Equation (\ref{eq:nT})
enable to derive the conditions for the downstream plasma to be
collisional.  For quantum plasma, the boundary $n=6.3 \times
10^{22}$~cm$^{-3}$ is equivalent to
\begin{equation}\label{eq:Quant}
n_0 =   6.3 \times 10^{22} \left( \frac{\hat{\gamma}-1}{\hat{\gamma}+1}\right)~~  \mathrm{cm}^{-3},
\end{equation}
where $\hat \gamma$ is the adiabatic index.
If the temperature in the downstream is low, $T < T_F$, the plasma in
the downstream region is degenerate for initial density $n_0$ larger
than this value.

For classical plasma, the boundary is determined by the condition
$k_BT = q^2 n^{1/3}$ (see Equation \ref{eq:EF}). Using Equation
(\ref{eq:nT}), this relation can be written as 
$(\gamma_0,n_0)$
%
\begin{eqnarray}\label{eq:Class}
m_e v_0^2 &=& q^2 \left( n_0 \frac{\hat{\gamma}+1}{\hat{\gamma}-1}
\right)^{1/3} \nonumber \\
\Rightarrow n_0 &=& N^* \left(\frac{\hat{\gamma}-1}{\hat{\gamma}+1}
\right)\beta_0^6,
\end{eqnarray}
where $N^* = 4.5 \times 10^{37}$~cm$^{-3}$ has been defined in
Equation (\ref{eq:N*}). The conditions set in Equations (\ref{eq:Quant},
\ref{eq:Class}) are equal at
\begin{equation}\label{eq:betac}
\beta_0= \left( \frac{8}{9 \pi^4} \right)^{1/6} \alpha = 0.0033 \ll 1,
\end{equation}
justifying the non-relativistic treatment.

\begin{figure}
\begin{center}
 \includegraphics[width=.9\textwidth]{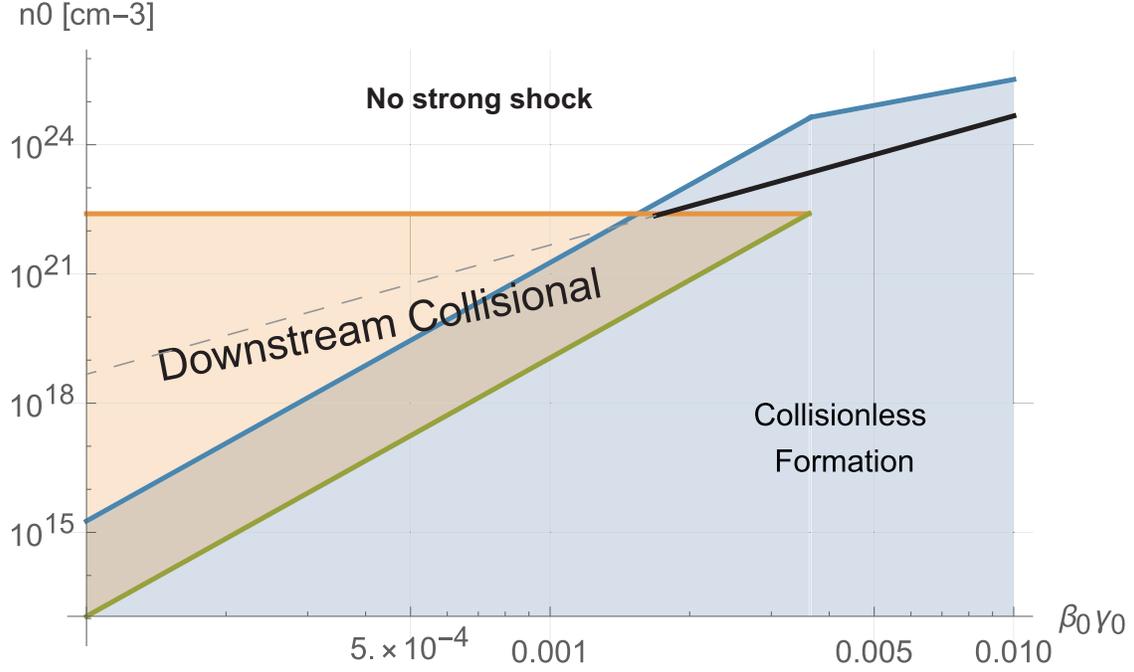}
\end{center}
\caption{If the pair plasmas shells have their initial momenta and
  densities $(\beta_0\gamma_0,n_0)$ located inside the pale orange
  triangle, the downstream is collisional and the resulting shock
  cannot accelerate particles. In producing this plot, we consider
  $\hat{\gamma}=5/3$. The bold black line pictures the no strong shock
  condition at $T=0$ discussed in section \ref{sec:T=0}.  At $T <
  T_F$, no strong shock forms above this line, for the speed of sound becomes
  larger than the collision speed.}\label{fig:COOL}
\end{figure}

The boundaries defined by Equations (\ref{eq:Quant}, \ref{eq:Class})
are shown in Figure \ref{fig:COOL}, together with the criteria for
collisional or collisionless shock formation. If the initial shells
are located inside the pale orange triangle, the downstream of the
resulting shock is collisional (fluid); outside of this regime, it is
collisionless.

Noteworthily, there is a region of the phase space in which the shock
formation is mediated by collisionless instabilities, while the
resulting shock has its downstream collisional. This region is
enclosed between the blue and the green lines in Figure
\ref{fig:COOL}. Comparing Equations (\ref{eq:F}), (\ref{eq:ratio1})
and (\ref{eq:Class}), one sees that these two lines are almost exactly
parallel, and are separated by a factor
\begin{equation}
\label{eq:factor}
\frac{\mathrm{Blue~frontier}}{\mathrm{Green~frontier}} \sim
\frac{128}{\pi}\frac{\hat{\gamma}+1}{\hat{\gamma}-1} = 163 ~~
\mathrm{for} ~~ \hat{\gamma}=5/3.
\end{equation}

The upper bounds of the weakly coupled domain displayed on Figure
\ref{fig:classi} translate to upper-bounds on Figure
\ref{fig:COOL}. If the colliding shells have initially either $\beta_0
> 0.0033$ ($v_0>1003$ km/s) or $n_0 > 1.57 \times 10^{22}$ cm$^{-3}$
(for $\hat{\gamma}=5/3$), then the downstream region of the resulting
shock is always collisionless. In both cases, particle acceleration
can occur.

\section{Electron/proton plasmas encounters}
\label{sec:case_ep}

We now adapt the previous results for the case of proton/electron
plasmas. The overall dynamic of the system in this scenario is
determined by the interaction of the protons. Following the structure
of the preceding sections, we first assess the binary collision
frequency before turning to the maximum growth-rate.

\subsection{Inter-shells collision frequency}
\label{sec:collfreqp}

The classical impact parameter for close binary Coulomb collisions in
electron-proton plasma reads $b_C = q^2/(\gamma_r m_p v_r^2)$,
%
%
where $m_p$ is the proton mass (see Equation \ref{eq:bC}). The quantum
impact parameter in this scenario is $b_Q = \hbar/p = \hbar/(\gamma_r
v_r m_p)$ (see Equation \ref{eq:bQ}).
%
%
The close collision frequency between protons of two different shells
is therefore
\begin{equation}\label{eq:coll_freqp}
  \nu_{ss} = n_0 v_r \pi b^2 = n_0 \frac{2\beta_0}{1+\beta_0^2}c ~
  \pi \max (b_C, b_Q)^2.
\end{equation}
One thus finds that in this scenario as well, the equality $b_C=b_Q$
is reached for $\beta_r = \alpha \sim 1/137$.

\subsection{Growth-rate for the collisionless interaction}
\label{sec:Taux_ep}

In the collisionless regime, the counter-streaming electrons first
turn unstable as the shells start overlapping. Once the electronic
instability has saturated, the counter-streaming protons turn unstable
\cite{Bret2015ApJL}. In the relativistic regime, the most unstable
mode of the counter-streaming protons over the bath of electrons is
still the Weibel instability, with a maximum growth-rate given
by \cite{Shaisultanov2012}
\begin{equation}\label{eq:GR_ep_R}
  \delta = 2 \frac{\beta_0}{\sqrt{\gamma_0}} \omega_{pp}.
\end{equation}
This result is identical to that in Equation (\ref{eq:gr_relat}),
after replacing the electron plasma frequency $\omega_{pe}$ by the
proton plasma frequency $\omega_{pp}^2=4 \pi n_0 q^2/m_p$.

In the non-relativistic regime, the same pattern occurs. Electrons are
stopped first before the counter-streaming protons become unstable
over the bath of electrons. In the limit of small velocity, $\beta_0
\ll 1$, the temperature of this electron bath approaches zero since its
thermal energy originates from the initial kinetic energy of the electron
beams. The fastest growing mode is found with a $\mathbf{k}$ aligned
with the flow \cite{BretPoPReview}. The dispersion equation for the
interaction is derived in Appendix \ref{app:A} and reads,
\begin{equation}\label{eq:NRion}
  \frac{2}{x^2}+\frac{R}{(Z-x)^2}+\frac{R}{(x+Z)^2}=1,
\end{equation}
where $x=\omega/\omega_{pe}$, $Z=kv_0/\omega_{pe}$ and $R=m_e/m_p$ is
the mass ratio (note that the frequency is measured in units of the
\emph{electron} plasma frequency).
This equation is solved in Appendix \ref{app:A}, yielding the maximum
growth-rate for this regime,
\begin{equation}\label{eq:NRionOK}
  \delta \sim \frac{\sqrt{3}}{2^{7/6}}R^{1/3} \omega_{pe} =
  \frac{\sqrt{3}}{2^{7/6}} R^{-1/6} \omega_{pp}.
\end{equation}
%
%
%

\begin{figure}
\begin{center}
\includegraphics[width=.7\textwidth]{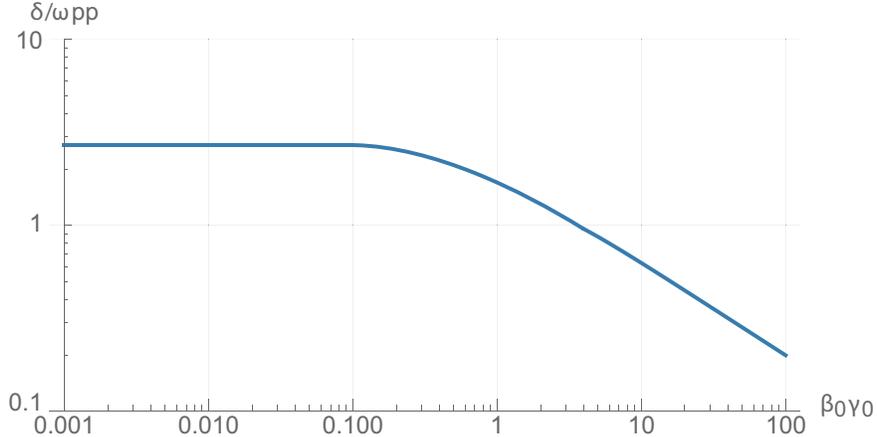}
\end{center}
\caption{Growth-rate of the fastest growing unstable mode in terms of
  $\beta_0\gamma_0$ for electron/proton plasmas interaction. In
  producing this plot, we took the mass ratio is $R = m_e/m_p =
  1/1836$. We interpolated between the non-relativistic (Equation
  \ref{eq:NRionOK}) and the relativistic (Equation \ref{eq:GR_ep_R})
  regimes.}
\label{fig:deltapi}
\end{figure}

\begin{figure}
\begin{center}
   \includegraphics[width=.7\textwidth]{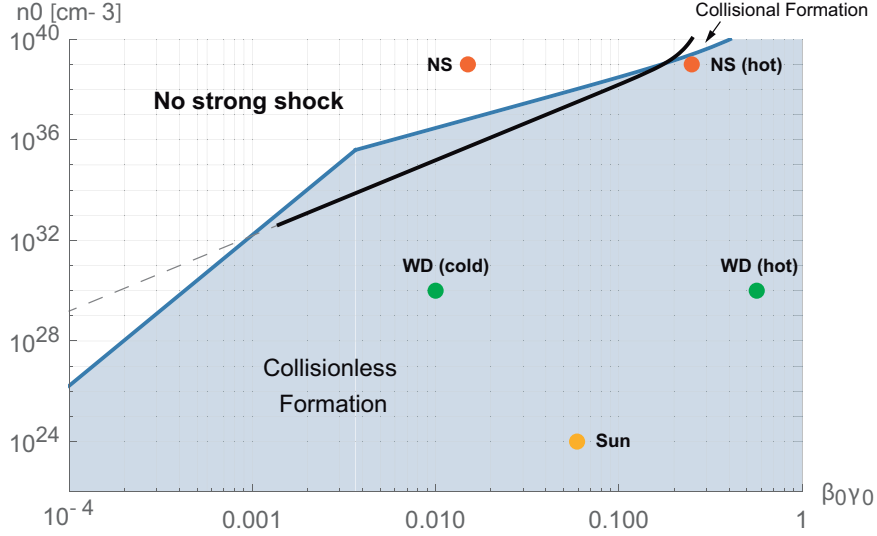}
  \end{center}
\caption{Frontier between the collisional and the collisionless
  domains for shock formation in the case of electron/proton
  plasmas. The colored regions pertain to astrophysical scenarios
  discussed in Section \ref{sec:obs}. The bold black line pictures the
  no strong shock condition discussed in section \ref{sec:T=0}.  At $T <
  T_F$, no strong shock forms above this line, for the speed of sound becomes
  larger than the collision speed.}
\label{fig:frontier_ep}
\end{figure}

In both the relativistic (Equation \ref{eq:GR_ep_R}) and
non-relativistic (Equations \ref{eq:NRionOK}) regimes, the growth rate
$\delta$ is linear in $\omega_{pp}$, namely it admits the form $\delta
= X \omega_{pp}$.  In order to fill the gap between these regimes, we
have implemented a simple first-order interpolation scheme. We point
out that a more accurate fluid model attributing to the electron bath
a temperature $3k_BT \sim (\gamma_0-1)mc^2$ gives very similar
results. Nevertheless, we dim the presentation of the fluid model
unnecessary because of (1) its length, (2) the small amount of
additional precision it brings, (3) the secondary relevance of this
point with respect to the main theme of this work and (4), the fact
that the regime corresponding to this interpolation ($\beta_0\gamma_0
\sim 1$) pertains to densities larger than $10^{40}$ cm$^{-3}$ (see
Figure \ref{fig:frontier_ep}).

The results of the growth rate are presented in Figure
\ref{fig:deltapi}. When comparing to the growth-rate for pair plasma
in Figure \ref{fig:delta}, we find that the local extremum has been
lost. This feature comes from the main difference between the two
settings, namely that for the electron/proton plasma, the proton
Weibel instability grows over a bath of electrons.

\subsection{Comparing the collision rate $\nu_{ss}$ and the instability growth rate $\delta$}
\label{sec:compap}

We continue with the road-map of the first part, comparing the time
scales for collisional and collisionless interactions in the
electron/proton case. In this scenario, we identify four separate
regimes.
\begin{itemize}
\item
For $\beta_r < \alpha=1/137$, Equation (\ref{eq:coll_freqp}) with
$\max (b_C, b_Q)=b_C$ is compared with Equation
(\ref{eq:NRionOK}). One obtains
\begin{equation}
  \frac{\nu_{ss}}{\delta} \sim \frac{\sqrt{\pi/3}}{4\ 2^{5/6}} ~
  \sqrt{\frac{n_0}{N^*}} R^{5/3} ~ {(1+\beta_0^2)^3 \over \beta_0^3
    (2\gamma_0^2-1)^2},
\end{equation}
where $N^*$ has been defined in Equation (\ref{eq:N*}).
\item
For $\beta_r >\alpha=1/137$, yet $\beta_0$ still non relativistic,
Equation (\ref{eq:coll_freqp}) with $\max (b_C, b_Q)=b_Q$, is to be
compared with Equation (\ref{eq:NRionOK}). One obtains,
\begin{equation}
  \frac{\nu_{ss}}{\delta} \sim 2^{-5/6}\sqrt{\frac{\pi}{3}} ~
  \sqrt{\frac{n_0}{N^*_1}} R^{5/3} ~ \frac{(1+\beta_0^2)}{\beta_0 (2
    \gamma_0^2-1)^2}
\end{equation}
where $N^*_1$ has been defined in Equation (\ref{eq:N1*}).
\item
In the intermediate regime when $\beta_0$ approaches unity, the value
of the growth-rate has been interpolated (see Figure
\ref{fig:deltapi}), $\delta = X \omega_{pp}$. One chooses $\max (b_C,
b_Q)=b_Q$ since $\beta_r > \alpha$ to obtain
\begin{equation}
  \frac{\nu_{ss}}{\delta} \sim \frac{\sqrt{\pi}}{X} ~
  \sqrt{\frac{n_0}{N_1^*}} R^{3/2} ~
  \frac{\beta_0}{\left(\beta_0^2+1\right) \beta_r^2\gamma_r^2}.
\end{equation}
\item
In the relativistic regime, Equation (\ref{eq:coll_freqp}) with $\max
(b_C, b_Q)=b_Q$, is to be compared with Equation
(\ref{eq:GR_ep_R}). One obtains,
\begin{equation}
    \frac{\nu_{ss}}{\delta} \sim \frac{\sqrt{\pi}}{16} ~
    \sqrt{\frac{n_0}{N_1^*}}R^{3/2} ~ \frac{2(1+\beta_0^2)\sqrt{\gamma_0}}{\beta_0^2(2\gamma_0^{2}-1)^2}.
\end{equation}
\end{itemize}
The frontier between the collisional and the collisionless domains for
shock formation is displayed in Figure \ref{fig:frontier_ep}.

\subsection{Conditions on $n_0$ and $\gamma_0$ for a collisionless downstream}
\label{sec:down_collp}

The strongly coupled regime in the $(T,n)$ parameter space is similar
to the one shown in Figure \ref{fig:classi}. Replacing the electron
mass by the proton mass in the calculations yields a threshold density
for weakly coupled degenerate plasmas of
\begin{equation}\label{eq:ncritQp}
  n > \frac{8}{9\pi^4} \frac{m_p^3 q^6}{\hbar^6} =  3.88 \times 10^{32}~\mathrm{cm}^{-3}.
\end{equation}
Note that for classical plasmas, the weakly coupled regime still demands $k_B T > q^2 n^{1/3}$.

The calculations conducted in Section \ref{sec:nT} to determine
$n(n_0,\gamma_0)$ and $T(n_0,\gamma_0)$ are straightforwardly
adapted. Equation (\ref{eq:nT}) now read $k_B T = (1/2) m_p v_0^2$ and $n = (\hat \gamma + 1)/(\hat \gamma -1) n_0$,
where the $1/2$ factor in the first equation accounts for the fact
that the initial kinetic energy of the electrons is negligible
compared to that of the protons ($k_BT \sim \frac{1}{2}m_p v_0^2$).

The downstream of the formed shock will therefore be collisional if
\begin{equation}
n_0 < 3.88 \times 10^{32}~\frac{\hat{\gamma}-1}{\hat{\gamma}+1} ~ \mathrm{cm}^{-3},
\end{equation}
when degenerate, and
\begin{equation}
n_0 > \frac{1}{8} \left( \frac{\hat{\gamma}-1}{\hat{\gamma}+1} \right) ~ \left(
\frac{m_p c^2}{q^2} \right)^3 \beta_0^6 =
3.47 \times 10^{46}~\frac{\hat{\gamma}-1}{\hat{\gamma}+1}~ \beta_0^6,
\end{equation}
when classical.

\begin{figure}
\begin{center}
\includegraphics[width=.8\textwidth]{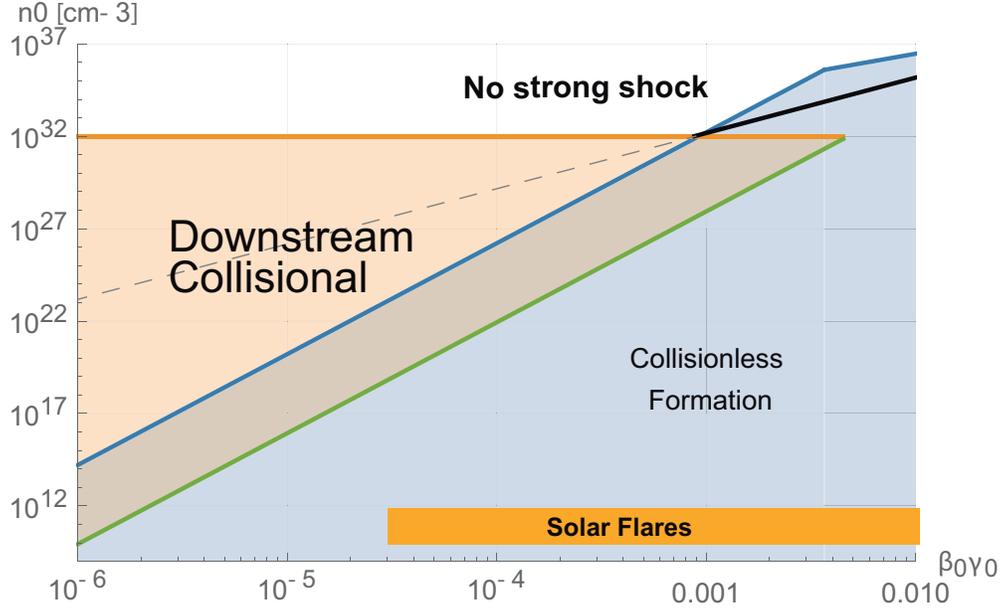}
  \end{center}
\caption{If the electron/proton plasma shells have their initial
  $(\beta_0\gamma_0,n_0)$ located inside the pale orange triangle, the
  downstream is collisional and the resulting shock cannot accelerate
  particles. We considered $\hat{\gamma}=5/3$. The colored region
  pertains to the solar flares scenario discussed in Section
  \ref{sec:obs}. The bold black line pictures the no strong shock condition
  discussed in section \ref{sec:T=0}.  At $T < T_F$, no strong shock forms
  above this line, for the speed of sound becomes larger than the
  collision speed.}\label{fig:COOLp}
 \end{figure}

The corresponding region is pictured in Figure \ref{fig:COOLp}. The
intersection of the two strongly coupled limits is found at the same
initial velocity as for the pair plasma, since the relevant value of $\beta_0$
(Equation \ref{eq:betac}) does not depend on the mass.

\section{About the small velocity spread regime: The ``no strong shock'' condition}
\label{sec:T=0}

In the calculations presented above, we neglected any velocity spread
$\Delta v$ within the shells, that is, we assumed $\Delta v \ll
v_r$. Noteworthily, the speed of
sound $c_s$ is proportional to $\Delta v$ (with a proportionality
constant of order unity). Formally, the Riemann problem consisting of two symmetric fluids colliding, always gives rise to 2 counter-propagating shock waves (see \cite{LandauFluid} p. 362, or \cite{Zeldovich} p. 89). But as shown in Appendix \ref{app:shock} by Eq. (\ref{eq:weak}), for $\mathcal{M}_0 \equiv v_0/c_s \ll 1$, the compression ratio of these shocks is $r \sim 1+ \mathcal{M}_0$. And since the distribution of accelerated particles is a power law of index $q \propto (r-1)^{-1}$ \cite{Blandford78}, such weak shocks will not be efficient particle accelerators.

Therefore, the equality $\Delta v \propto c_s=v_r$, which marks
 the limit of our calculations, also
indicates the limit beyond which no strong shock forms, hence particle acceleration does not occur. Let us now further assess this limit.

Cold plasmas having $T<T_F$ are degenerate (see Figure
\ref{fig:classi}). In this regime, the speed of sound is a function of
the Fermi temperature $T_F$, hence of the density. On the other hand,
for $T>T_F$ the colliding shells are classical and the speed of sound
becomes a function of the temperature.  We shall now examine the limit
$c_s=v_r$ first for degenerate plasma, and then in the classical
regime.

\subsection{``No strong shock'' condition for cold, degenerate plasma}
Since the Fermi temperature increases with the density, for a given
shell encounter speed $v_r$ the condition for strong shock formation, $v_r >
c_s$ implies the existence of a critical (maximal) density,
$n_{0,ns}$ \footnote{The subscript ``ns'' in $n_{0,ns}$ stands for
  ``no shock''.} beyond which the encounter cannot produce a strong shock.
At higher densities $n>n_{0,ns}$, the Fermi temperature $T_F$, hence the
speed of sound, is larger than the relative speed of the collision.

In order to find the no strong shock condition, one needs to discriminate
between the strongly coupled and the weakly coupled regimes. In the
weakly coupled regime, simple analytical expressions exist in the
non-relativistic (Newtonian) and the ultra-relativistic limits.

\subsubsection{Weakly coupled plasma regime}

The general expressions for the energy density and the pressure in a
degenerate Fermi gas are given by \cite{LandauStat},
\begin{equation}
u ={c \over 8 \pi^2 \hbar^3} \left\{ p_F \left( 2 p_F^2 +
m^2 c^2 \right)\sqrt{p_F^2 + m^2 c^2} - \left( m c \right)^4
\sinh^{-1} \left({p_F \over m c} \right) \right\},
\label{eq:u}
\end{equation}
\begin{equation}
P ={c \over 8 \pi^2 \hbar^3} \left\{ p_F \left( {2 \over 3} p_F^2 -
m^2 c^2 \right)\sqrt{p_F^2 + m^2 c^2} + \left( m c \right)^4
\sinh^{-1} \left({p_F \over m c} \right) \right\},
\label{eq:p}
\end{equation}
where
\begin{equation}\label{eq:pF}
p_F = (3 \pi^2)^{1/3} n^{1/3} \hbar
\end{equation}
is the Fermi momentum, and
$m$ is the particle mass (electron or proton).

Simple expressions which enable exact analytical calculations exist in
both the Newtonian and the ultra-relativistic limits. In the Newtonian
limit, $p_F \ll m c$, Equations (\ref{eq:u}, \ref{eq:p}) are
approximated by
\begin{equation}
\label{eq:P_weak_NR}
u(N.R.) \simeq \frac{3  \left(3 \pi ^2\right)^{2/3}}{10}
\frac{\hbar^2}{m} n^{5/3},~~P(N.R.) = \frac{2}{3} u(N.R.).
\end{equation}
On the other hand, in the ultra-relativistic limit, $p_F \gg m c$ one
obtains \cite{Chandra1931},
\begin{equation}
\label{eq:PChandra}
u (rel.) \simeq \frac{3}{8}\left( \frac{3}{\pi} \right)^{1/3} h c ~
n^{4/3},~~P(rel.) = {u(rel.) \over 3}.
\end{equation}
The general expression for the speed of sound in a degenerate Fermi
gas, which is correct in both the Newtonian and relativistic regimes
is $c_s^2 = c^2 {dP\over d \bar u}$, where $\bar u = n m c^2 + u$. In
the Newtonian regime, $u \ll n m c^2$, this reduces to $c_s^2 =
p_F^2/3m^2$, while in the ultra-relativistic regime $c_s^2 = c^2/3$,
similar to the well-known result obtained for classical gas.

In order to find the criteria for strong shock formation, it is easier to work
in the comoving frame of one of the colliding shells. The comoving
density $n$ is related to the lab-frame density $n_0$ by
$n=n_0/\gamma_0$. In this frame the second shell is approaching at
velocity $v_r$, given by Equation (\ref{eq:vrell}). Equating this
velocity with the speed of sound, $c_s$, one finds the maximal density
that allows formation of strong shocks. In the Newtonian regime this density is
%
%
\begin{equation}\label{eq:n0_ns_nr}
n_{0,ns} = \frac{8 \sqrt{3}}{\pi ^2} \left( \frac{m c}{\hbar} \right)^3 \frac{\beta_0^3}{(1+\beta_0^2)^3}\gamma_0.
\end{equation}
In the relativistic regime, equating the colliding speed $v_r$ with the
sound speed $c_s = c/\sqrt{3}$ results in an asymptotic value of $\beta_0$,

\begin{equation}\label{eq:beta_0crit}
  \beta_0 = \sqrt{3} - \sqrt{2} \sim 0.32 ~~\Rightarrow ~~
  \beta_0\gamma_0 = \frac{1}{2}\sqrt{\sqrt{6}-2} \sim 0.33.
\end{equation}
We thus conclude that at faster shells velocities encounter, strong shock
waves will always form.

\subsubsection{Strongly coupled plasma regime}
At sufficiently low temperatures and low densities, namely $n_0 < 6.3
\times 10^{22}~{\rm cm}^{-3}$ (i.e. Fermi energy $E_F < q^2n^{1/3}$)
the plasma particles are strongly coupled. In this regime, the Fermi
energy is lower than the Coulomb potential. To the best of our
knowledge, there is no simple expression for the speed of sound
in this regime, which comprises warm dense matter (solid density at eV temperatures - see \cite{Fortney2010} and references therein) and condensed matter. These limitations, though, do not substantially limit the breath of the present work, because shocks formed in such regimes are definitively collisional, hence clearly poor particles accelerators.
We therefore represent the no strong shock boundary in this regime using dashed lines in Figures \ref{fig:ncrit}, \ref{fig:COOL}, \ref{fig:frontier_ep} and \ref{fig:COOLp}.

\bigskip

The no strong shock criteria is plotted in thick black lines in Figures
\ref{fig:ncrit} and \ref{fig:COOL} for the pair plasmas scenario. Strong
shocks cannot form above these lines. In the case of electron/proton
plasma, these calculations can be straightforwardly adapted replacing
the electron mass by the proton mass. The corresponding criteria is
plotted by thick black lines in Figures \ref{fig:frontier_ep} and
\ref{fig:COOLp}.

\subsection{``No strong shock'' condition in the classical plasma regime}

As long as $T<T_F(n)$, the shells can be regarded as cold, and the
degenerate results at T=0 apply. However, as sufficiently low
densities, $T>T_F(n)$ and the speed of sound varies with the temperature
rather than the density.

In this classical regime, the speed of sound is given by
$c_s = c \sqrt{(\partial P/\partial u)_s}$, where $u$ is the energy
density. Simple analytical expressions exist in the non-relativistic
($T \ll mc^2$) and relativistic ($T \gg mc^2$) limits:
\beq
c_s = \left\{ \ba{ll}
\sqrt{\hat \gamma {k_B T \over m}} & (T \ll m c^2), \\
{c \over \sqrt{3}} & (T \gg m c^2),
\ea \right.
\label{eq:c_s_classical}
\eeq
where $\hat \gamma $ is the adiabatic index.

Similar to the degenerate case, in order for a strong shock to form, the
requirement is $v_r > c_s$. Using Equation \ref{eq:vrell}, one finds
that for low, non-relativistic temperatures, the minimum velocity
scales as $\beta_0 \propto \sqrt{T}$. For hot (relativistic) plasmas,
the speed of sound is the same in both the classical and the
degenerate regimes, and therefore the minimum value of $\beta_0$ that
enables the production of strong shocks saturates to the value of $\beta_0$
given by Equation (\ref{eq:beta_0crit}).  This coincidence of the
degenerate and classical values of the minimum of $\beta_0$ allowed
for strong shock formation was to be expected, since both merge for $T=T_F$.

\begin{figure}
\begin{center}
\includegraphics[width=.8\textwidth]{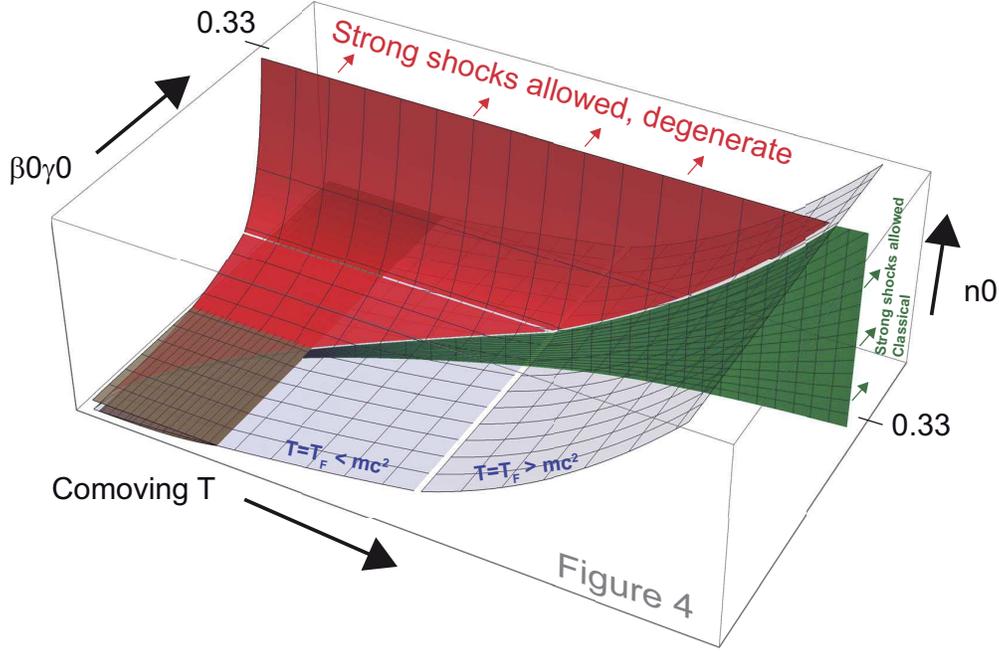}
  \end{center}
\caption{Sketch (not to scale) of the various domains involved in the
  article, in the phase space of parameters $(n_0, \beta_0, T)$. Strong shock
  formation is only allowed for systems located behind the red
  surface, or behind the green surface. Only weak shocks are formed in the rest
  of the parameter space, because the speed of sound exceeds the speed
  of the collision. Figure \ref{fig:classi} corresponds to a 2D cut at
  $\beta_0\gamma_0=0$, of this 3D plot.}
\label{fig:3D}
 \end{figure}

\subsection{Complete parameter space region for strong shock formation}
The results of this section are summarized in Figure \ref{fig:3D}.
This figure generalizes the results presented in Figures
\ref{fig:ncrit}, \ref{fig:COOL}, \ref{fig:frontier_ep} and
\ref{fig:COOLp} by adding a third dimension, namely, the comoving
shells temperature.

The purple surface pictures systems with $T=T_F$. Above it, the plasma
is degenerate, while below it is classical. As is evident, the slope
breaks at $k_B T = m c^2$, as at lower temperatures the plasma is
non-relativistic and $T_F \propto n^{2/3}$, while in the relativistic
limit $T_F \propto n^{1/3} \rightarrow n \propto T_F^3$. This purple
surface divides the parameter space into two regimes: above this
surface, $T<T_F$ and the plasma is degenerate, while below this
surface, $T>T_F$ and the plasma is classical.

The two brown surfaces represent the $k_BT=q^2n_0^{1/3}$ and
$k_BT_F=q^2n_0^{1/3}$ limits. Systems located between these two
surfaces are strongly coupled. These surfaces therefore represent a
3-dimensional extension of the results presented in Figure
\ref{fig:classi} (red lines).

The red/green surfaces picture the ``no strong shock'' condition ($v_r<c_s$)
in the degenerate/classical plasmas, respectively. In both regimes, as
long as the plasma is in the Newtonian regime, increasing the
temperature requires an increase in the collision velocity $\beta_0$
as $\beta_0 \propto T^{1/2}$ in order to enable the formation of
strong shocks. Thus, strong shocks can only be formed ``outside'' of the pictured
surfaces.

As the density increases, the Fermi temperature becomes relativistic,
$T_F > m c^2$, and the condition on the collision velocity saturates at $\gamma_0 \beta_0 \simeq 0.33$ (see Equation
\ref{eq:beta_0crit}). At higher collision speeds, strong shock could always
form, regardless of the initial densities, the plasma temperatures or
the nature of the plasma, being either classical or degenerate.



\section{Astrophysical Implications}
\label{sec:obs}

A key difference between collisional and collisionless shocks rely on
the (theoretical) ability of the later to both accelerate particles to
high energies as well as generate strong magnetic fields. These
phenomena cannot occur in collisional shocks, due to the rapid
thermalization of the plasma in the downstream region, which
suppresses the growth of any seed magnetic fields, as well as
thermalize energetic particles \cite[e.g.][]{Longair11}.

Relativistic outflows in the form of astronomical jets are most easily
observed at the exterior of many astronomical objects, ranging from
supernova, X-ray emitting binaries (neutron stars or black holes)
\cite{Fender06, Fender16}, gamma-ray bursts (GRBs) \cite{Meszaros06,
  Peer15} or supermassive black holes in active galactic nuclei (AGNs)
\cite{BBR84, UP95}. While these objects inevitably involve shock waves,
the low environmental densities imply that these shocks are formed
collisionless, and so are their downstream regions.

The situation is somewhat more complicated in the interior of the
various astronomical objects. We plot in Figure \ref{fig:frontier_ep}
typical values of the interior density as well as the normalized
velocities associated with random fluctuations of different
objects. The velocities are derived based on temperature estimate in
the interior of the different objects, considering that forming a
shock requires motions faster than the thermal velocity.  We consider
the solar interior, cooled white dwarfs (WD) (whose temperature vary
in the range $3 \times 10^5~-10^9~{}^\circ$K) \cite{ST83}, and neutron
stars (NS), having characteristic temperatures of $0.1 - 30$~MeV
\cite{LP07}. While shock waves that are generated in the interior of
main sequence stars and WDs are always in the collisionless regime,
shocks that can potentially form in the interior of NS can in
principle be generated in the collisional regime, though they may fall
into the ``no shock'' parameter space region. As these shocks
propagate outwards, they will eventually propagate towards lower
density and potentially higher velocities regime, and may therefore
move into the collisionless regime of the parameter space.

Pair dominated plasmas are expected to occur in few astronomical
scenarios.  One of the widely discussed scenarios is that of
pair-instability supernovae. Massive stars ($M\gtrsim 100 M_\odot$) form
large helium cores that reach carbon ignition with masses in excess of
$\sim 45 M_\odot$. After helium burning, cores of this mass will
encounter the electron-positron pair instability, collapse and ignite
oxygen and silicon burning explosively. If explosive oxygen burning
provides enough energy, the result is a ``pair-instability supernovae''
\cite{BAC84, HW02}. In recent years, there were several observational
evidence for this mechanism, e.g., in SN2006gy \cite{Smith+07} or
SN2007bi \cite{GalYam+09}. In Figure \ref{fig:ncrit}, we plot possible
parameter space values for this scenario.

A second scenario is that of gamma-ray bursts (GRBs). The leading
model to explain the observed variable light-curves in these objects,
the GRB ``fireball'' model \cite{RM94} invokes internal shocks. These
shocks can in principle occur at radii as small as a few times the
Schwarzschild radius of $\sim 10~M_\odot$ black hole, namely at
$\gtrsim 10^7$~cm. The densities are similar to that at the interior
of massive stars. As these shocks occur below the photosphere, a
significant number of pairs are created
\cite{Beloborodov17}. Equilibration between pair production and
annihilation results in density ratio of $n_\pm/n_e \sim 10$
\cite{PW04}. The typical values of density and velocity in this
scenario is similar to that of pair-instability supernovae, and is
similarly plotted in Figure \ref{fig:ncrit}.

We thus conclude that in these objects, the shock waves can initially
be formed as collisional, though as they propagate outward, they
become collisionless. As a consequence, the time available for particle
acceleration and magnetic field generation in these shock waves could be
limited.

Another environment of interest is that of solar flares. With typical
electron densities of $\sim 10^{10} - 10^{12}~{\rm cm^{-3}}$
\cite{Asc02} and typical velocities in the range $30- 10^4~{\rm km/s}$
\cite{Asc02, JFL17}, the resulting shock waves are expected to be
formed collisionless, but the downstream region of the low velocity
shocks (at least, initially) could potentially be collisional. While
the flares will eventually propagate into lower density region, these
conditions limit their ability to accelerate particles at their
initial phases. The corresponding parameters space values are plotted on Figure \ref{fig:COOLp}.

\section{Conclusions}
\label{sec:conclusions}

This article deals with the properties of the shock waves that are
formed when two symmetrical plasmas run into each other, and in
particular in the question of collisionality. We assessed the answers to
two questions: 1/ When is the interaction mediated by close Coulomb
binary collisions or collective plasma instabilities? and 2/ Once a
shock has been formed, when is its downstream collisionless?

The answer to the first question is given in Figure \ref{fig:ncrit}
for the pair plasmas and Figure \ref{fig:frontier_ep} for the
electron-proton plasmas. The answer to the second question is
presented in Figures \ref{fig:COOL} \& \ref{fig:COOLp} for pair and
electron-proton plasmas respectively.

As we showed here, the switch from collisional to collisionless regime
bears consequences on the time scale of the shock formation. In the
collisionless regime, the shock formation time is determined by the
growth-rate of the unstable interaction between the two shells. In the
collisional regime, on the other hand, the shock formation time is set
by the frequency of close binary collisions.

Moreover, a shock which formation not mediated by collisionless
plasma instabilities will not inherit the downstream electromagnetic
patterns formed by these instabilities. Collisions of shells with
curved boundaries may trigger downstream vorticity that could generate
magnetic fields \cite{bond2017}. But colliding planar shells like
those considered here, will yield a field-free downstream if the
formation is collisional.

Understanding the properties of the downstream region (the second
question outlined above) has the important consequence of determining
the ability of the downstream region to accelerate particles. As
discussed above, a collisional downstream leads to a suppression of
particle acceleration. For two plasma shells initially located in the
orange triangle plotted in Figure \ref{fig:COOL} for pair plasmas, or
Figure \ref{fig:COOLp} for electron/proton plasmas, the resulting
shock will not be able to accelerate particles. As a result, the
radiative signature of the system would be dramatically different than
that of a system in which the downstream is collisionless.

In Section \ref{sec:obs}, we discussed several astrophysical settings
where a shock may be formed as collisional, and, during its
propagation inside the object, its properties will be modified from a
collisional to a collisionless medium. The key consequences of this
transition is the limiting ability of these shocks to accelerate
particles to high energies and to generate magnetic fields. This
implies stringent constraints on the abilities of these objects to be
the sources of high energy cosmic rays, as well as modification of the
resulting spectra. We leave a detailed study of the spectra expected
under various conditions to future work.  Similar situations can be
found in the context of Inertial Confinement Fusion
\cite{CollLessICF2017}. While the medium considered here are spatially
homogeneous, it would be worth studying how a shock transiting from one
kind of medium to another, evolves.

In this work, we focused on ``classical'' shocks, namely shocks that
are mediated either by collective plasma effects or by Coulomb
collisions. If shock waves occur in regions of high optical depth in
which the mean free path is smaller than the shock size, they may be
mediated by photons scattered back and forth the upstream and
downstream regions \cite{Budnik+10, Beloborodov17, LBV17},  provided that the radiation energy density dominates the energy density of the gas. This
scenario is further expected to modify the shock properties; in
particular, no particle acceleration is expected. We leave a full
treatment of the properties of these radiative-mediated shocks to a
future work.  We do point out, though, that in all astrophysical implications considered in this work, namely neutron stars, white dwarfs and stellar interiors, radiative pressure is sub-dominant over the gas pressure, and thus the formed shock waves are not expected to be radiatively-mediated. Radiative-mediated shocks are expected in stellar envelopes, stellar winds, supernov{\ae} explosions, and possibly in gamma-ray bursts.

Encounter of partially ionized shells could be worthy of
investigation. In such a setting, the outcome may sharply depend on
the intra-shell coupling between the ionized and the neutral
particles. If intra shells collisions are rare, both component may act
separately. The ionized part should form a collisionless shock on a
time scale given by plasma instabilities. Meanwhile, the neutral
component could form a shock on a time scale defined by the collision
frequency between neutrals of different shells. But if both components
are coupled, then the first one to form a shock may drag the other
into the formation of a single, common, shock. Such scenarios will be
studied in future works.

Future works could also focus on studying temperature effects. The present
calculations are valid as long as the shells' velocity spread $\Delta
v$ fulfills $\Delta v \ll v_r$. Larger spreads should affect the
binary collisions frequencies, the maximum growth-rates, and also the
Rankine-Hugoniot (RH) conditions used to determine the collisionality
of the downstream. While such effects on the binary collisions
frequencies, or on the RH conditions are accessible, there are so far
no analytical formulas available for the maximum growth-rate in terms
of the temperature and $\gamma_0$ \cite{BretPRE2010}. Such progresses
are therefore a prerequisite before one can elaborate on larger
temperature effects. Note however that as discussed in section
\ref{sec:T=0}, no strong shock should form when $\Delta v > v_r$,
since the speed of sound becomes larger than the collision speed in
this limit. Therefore, as far as the velocity spread is concerned,
this work investigates the case $\Delta v/v_r \ll 1$, and the regime
left to explore is simply $\Delta v/v_r$ of order unity.

\section{Acknowledgments}
A.B. acknowledges support by grant ENE2016-75703-R from the Spanish
Ministerio de Educaci\'{o}n. A.P. acknowledges support by the European
Union Seventh Framework Program (FP7/2007-2013) under grant agreement
no. 618499, and support from NASA under grant
no. NNX12AO83G. A.B. thanks Gustavo Wouchuk, Roberto Piriz, Rony Keppens, Rolf Walder and Lorenzo Sironi for
fruitful discussions.

\appendix

\section{Determination of $n(n_0,\gamma_0)$ and $T(n_0,\gamma_0)$}
\label{sec:nT}

The fact that the collisional regime lies deep within the
non-relativistic domain as is shown in Figure \ref{fig:classi},
suggests that a non-relativistic treatment is appropriate. Once the
shock is formed, the shock frame is well defined. In this frame, we
use the subscripts ``1'' (``2'') to describe upstream (downstream)
quantities.  In the non-relativistic regime and for zero upstream
pressure ($P_1=0$), the Rankine-Hugoniot (RH) relations read
\cite{Zeldovich},
\begin{eqnarray}\label{eq:RHn}
  \frac{n_2}{n_1}&=&\frac{\rho_2}{\rho_1}=\frac{\hat{\gamma}+1}{\hat{\gamma}-1},
  \\
\frac{v_2}{v_1}&=&\frac{\hat{\gamma}-1}{\hat{\gamma}+1}.
  \nonumber
\end{eqnarray}
Here, $\hat{\gamma}$ is the adiabatic index of the gas, the $\rho_1,
\rho_2, v_1, v_2$ are the mass densities, upstream and downstream
velocities respectively. The RH relation for the downstream pressure
$P_2$ leads to
\begin{eqnarray}\label{eq:RH_P}
  P_2 &=& \rho_1 v_1^2 - \rho_2 v_2^2 \nonumber \\ &=& \rho_1 v_1^2 -
  \rho_1 \frac{\hat{\gamma}+1}{\hat{\gamma}-1} \left( v_1
  \frac{\hat{\gamma}-1}{\hat{\gamma}+1} \right)^2 \nonumber\\ &=&
  \rho_1 v_1^2 \frac{2}{\hat{\gamma}+1}.
\end{eqnarray}
Since the plasma is non relativistic, the upstream velocity as
written in the downstream frame is simply $\beta_0=v_0/c$. One thus
have,
\begin{equation}
  v_0 = v_1 - v_2 = v_1 - \frac{\hat{\gamma}-1}{\hat{\gamma}+1} v_1 =
  \frac{2}{\hat{\gamma}+1} v_1.
\end{equation}
Using this result in Equation (\ref{eq:RH_P}) then gives,
\begin{eqnarray}
  P_2 &=& \rho_1 v_1^2 \frac{2}{\hat{\gamma}+1} = \rho_1 \left(
  \frac{\hat{\gamma}+1}{2} v_0 \right)^2 \frac{2}{\hat{\gamma}+1}
  \nonumber \\ &=& \rho_0 \frac{\hat{\gamma}+1}{2} v_0^2= 2 n_0 m_e
  \frac{\hat{\gamma}+1}{2} v_0^2,
\end{eqnarray}
where $\rho_1$ has been replaced by $\rho_0$ in the last line, as no
relativistic boosting exists, while the factor 2 comes because there
are $n_0$ electrons and $n_0$ positrons per unit volume.

Denoting the energy density by $u$, the downstream pressure $P_2$ reads,
\begin{equation}
P_2 = (\hat{\gamma}-1) \rho u = (\hat{\gamma}-1) n k_B T.
\end{equation}
One finally obtains,
\begin{equation}
n k_B T = {m_e} \frac{\hat{\gamma}+1}{\hat{\gamma}-1} n_0 v_0^2,
\end{equation}
namely
\begin{eqnarray}\label{eq:nT}
k_B T &=& {m_e} v_0^2, \nonumber \\
n   &=& n_0 \frac{\hat{\gamma}+1}{\hat{\gamma}-1},
\end{eqnarray}
which are the relations we needed. The first one reads $k_BT =
\frac{1}{2}\left(2m_e v_0^2\right)$, which simply states that the downstream
thermal energy originates from the upstream kinetic energy. 

\section{Derivation of the dispersion equation (\ref{eq:NRion})}\label{app:A}
Suppose we have $a \in \mathbb{N}$ cold beams made of species of densities $n_j$, masses $m_j$ and velocities $\mathbf{v}_j = v_j \mathbf{e}_z$, the system being overall charge and current neutral. The dispersion equation for longitudinal waves with $\mathbf{k} \parallel \mathbf{e}_z$ reads (see \cite{Ichimaru}, p. 137),
\begin{equation}\label{eq:DGal}
  \sum_{j=1}^a \frac{\omega_{pj}^2}{(\omega - \mathbf{k} \cdot \mathbf{v}_j )^2} = 1,
\end{equation}
where $\omega_{pj}^2 = 4\pi n_j q^2/m_j$ is the plasma frequency of specie $j$. The unstable system considered in Section \ref{sec:Taux_ep} accounts for 3 species: the 2 counter-streaming proton beams, and the cold electronic background. With the dimensionless variables used, (\ref{eq:DGal}) then gives,
\begin{equation}\label{eq:D(x,Z)}
 D(x,Z) \equiv \frac{2}{x^2}+\frac{R}{(Z-x)^2}+\frac{R}{(x+Z)^2}-1=0,
\end{equation}
which is Equation (\ref{eq:NRion}).

This dispersion equation can be derived either using a multiple cold
fluids model \cite{califano3}, or taking the kinetic dispersion
equation for such waves, and considering a distribution function which
is the sum of Dirac delta functions (like in \cite{Ichimaru}).

  \begin{figure}
  \begin{center}
\includegraphics[width=.45\textwidth]{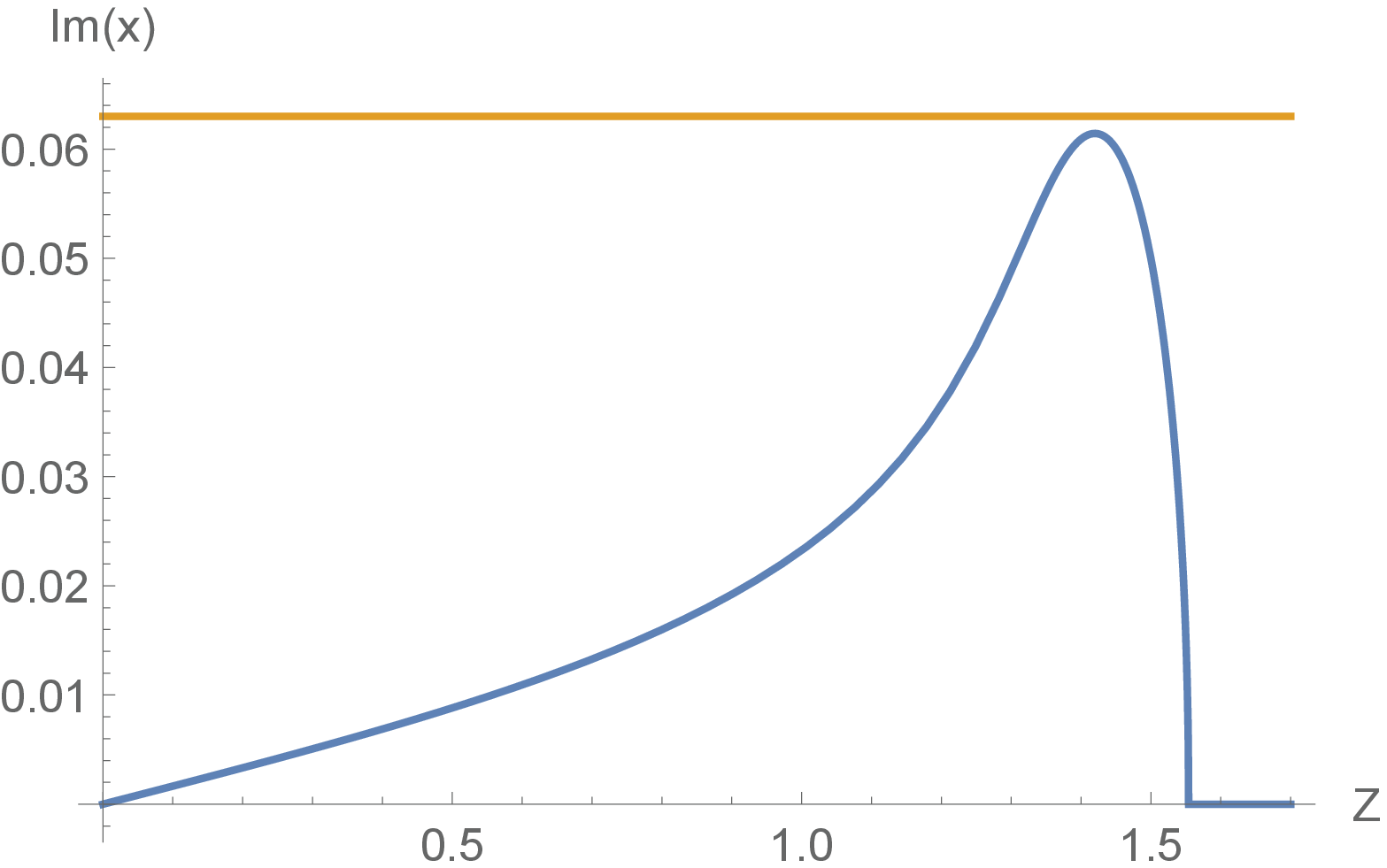}\includegraphics[width=.45\textwidth]{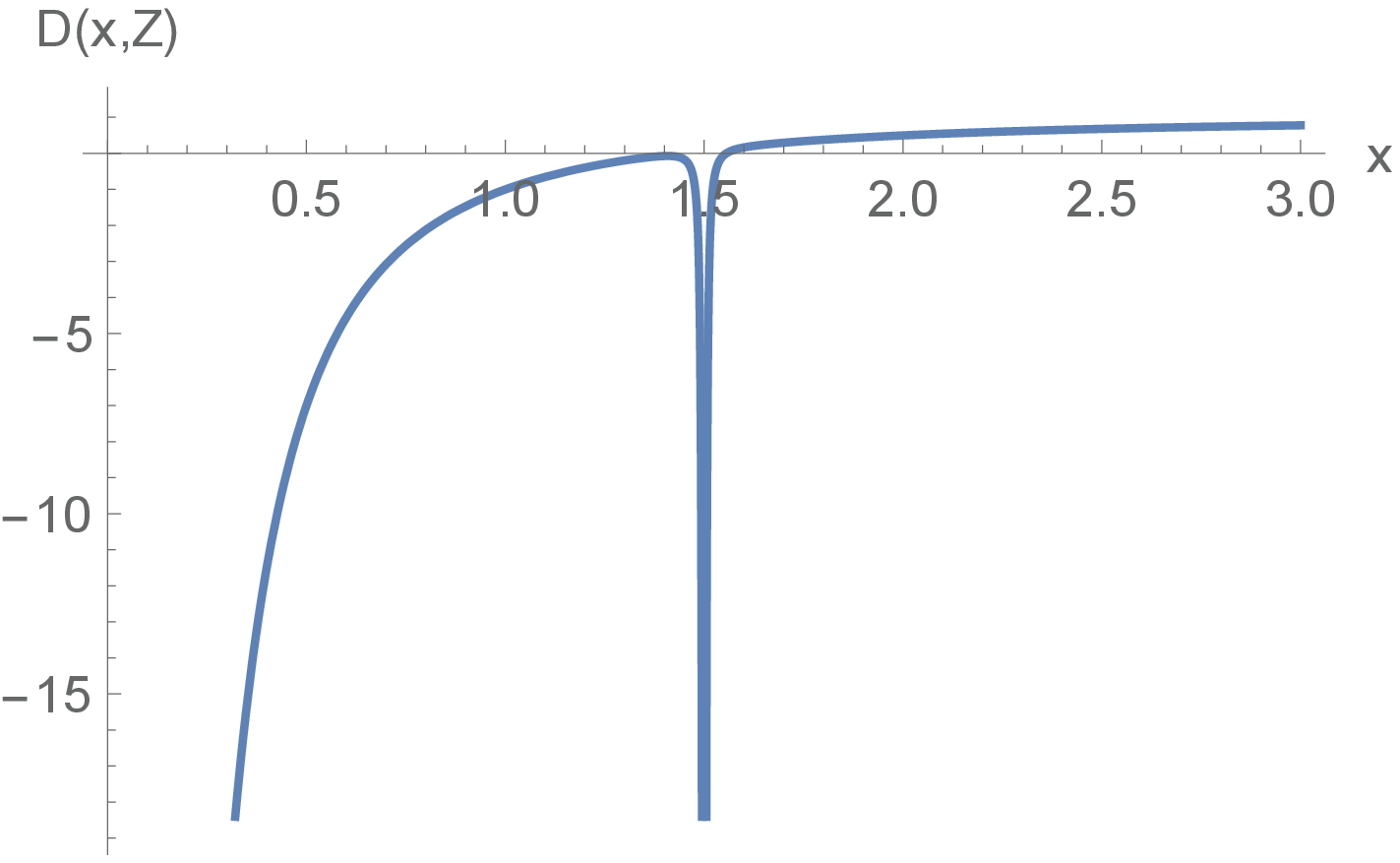}
  \end{center}
  \caption{\textbf{LEFT}: growth-rate given by Im$(x)$ with
    $x/D(x,Z)=0$ for $R=1/1836$. The orange line shows the result
    given by Equation (\ref{eq:NRionOK}). \textbf{RIGHT}: Plot of the
    dispersion function $D(x,Z)$ defined in Equation
    (\ref{eq:D(x,Z)}), for $R=1/1836$ and $Z=1.5$.}\label{fig:D(x,Z)}
 \end{figure}

The growth-rate Im$(x)$, with $x/D(x,Z)=0$, is plotted on Figure \ref{fig:D(x,Z)}-Left for $R=1/1836$. It reaches a maximum near $Z = \sqrt{2}$. This can be understood noting that since $R \ll 1$, the roots of the equation have to be close to the roots of the same equation, but with $R=0$, that is $x=\pm\sqrt{2}$. It follows that for $0<R\ll 1$, the numerators of the terms $\propto R$ have to be small if these terms are to bring a significant contribution to the equation. This implies in turn $Z \sim \pm\sqrt{2}$.

The dispersion function $D(x,Z)$ is plotted on Figure \ref{fig:D(x,Z)}-Right for $x>0$, $R=1/1836$ and $Z=1.5$. It is an even function of $x$, which is the reason why only the $x>0$ part is plotted. Let us now focus on the root located near $x=\sqrt{2}$. It is primarily determined by the terms $2/x^2$ and $R/(x-Z)^2$ of the dispersion equation. We can therefore study the solution located near $x=\sqrt{2}$ by neglecting $R/(x+Z)^2$, that is, solving,
\begin{equation}\label{eq:disper_redu}
  \frac{2}{x^2}+\frac{R}{(Z-x)^2}=1.
\end{equation}
We can solve approximately this equation following a method derived long ago \cite{Bludman}. Knowing that the maximum growth-rate $\delta$ is found for $x \sim \sqrt{2}$, we write $x = \sqrt{2} + i\delta$, with $\delta \in \mathbb{R}$. We then assume $|Z-\sqrt{2}| \ll |\delta|$ (which is later verified) so that Equation (\ref{eq:disper_redu}) becomes,
\begin{equation}
  \frac{2}{(\sqrt{2} + i\delta)^2}-\frac{R}{\delta^2}=1.
\end{equation}
A Taylor expansion of the first term, and some straightforward
algebra, gives the growth-rate (\ref{eq:NRionOK}).

  \begin{figure}
  \begin{center}
   \includegraphics[width=\textwidth]{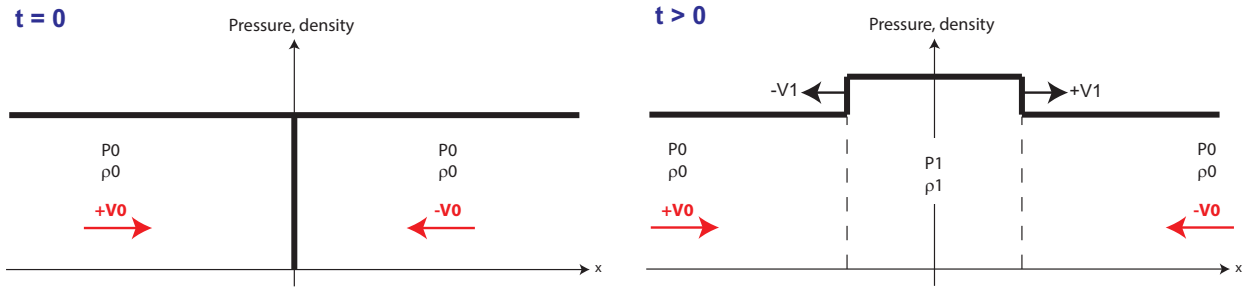}
     \end{center}
  \caption{Riemann problem considered.}\label{fig:riemann}
 \end{figure}

\section{Density jump of shocks formed a low collision velocity}\label{app:shock}
The Riemann problem pictured on Figure  \ref{fig:riemann} consists of two symmetric fluid colliding. It gives rise to 2 counter-propagating shock waves (see \cite{LandauFluid} p. 362, or \cite{Zeldovich} p. 89). Because the efficiency of particle acceleration depends on the compression ratio $r$, let us compute $\rho_1$ in terms of $v_0, P_0, \rho_0$.

In the reference frame of the shock front, the upstream comes at $V_0+V_1$, and the downstream goes at $V_1$. We thus have,
\begin{eqnarray}
  \rho_1V_1 &=& \rho_0(V_1+V_0), \\
  \rho_1 &=& \rho_0 \frac{(\hat{\gamma}+1)\left( \mathcal{M}_0 + M_1 \right)^2}{(\hat{\gamma}-1)\left( \mathcal{M}_0 + \mathcal{M}_1 \right)^2 + 2} ,
\end{eqnarray}
where the first equation is the conservation of matter, the second, the Rankine-Hugoniot jump condition for the density, and
\begin{eqnarray}
  \mathcal{M}_0 &=& \frac{V_0}{\sqrt{\hat{\gamma} P_0/\rho_0}}, \\
  \mathcal{M}_1 &=& \frac{V_1}{\sqrt{\hat{\gamma} P_0/\rho_0}}.
\end{eqnarray}
Setting $r = \rho_1/\rho_0$, these equations read,
\begin{eqnarray}
  r    &=& \frac{\mathcal{M}_0}{\mathcal{M}_1} + 1, \\
  r     &=&  \frac{(\hat{\gamma}+1)\left( \mathcal{M}_0 + \mathcal{M}_1 \right)^2}{(\hat{\gamma}-1)\left( \mathcal{M}_0 + \mathcal{M}_1 \right)^2 + 2}.
\end{eqnarray}
These expressions must be solved for $r$ and $\mathcal{M}_1$, under the constraint $\mathcal{M}_0+\mathcal{M}_1 > 1$.

Eliminating $r$ gives a third order polynomial which can be factored by $(\mathcal{M}_0 + \mathcal{M}_1)$. The remaining second order polynomial can be solved exactly. Its positive root is,
\begin{equation}
  \mathcal{M}_1 = \frac{1}{4} \left[ (\hat{\gamma}-3)\mathcal{M}_0 + \sqrt{(\hat{\gamma}+1)^2\mathcal{M}_0^2+16}\right].
\end{equation}
We have $\mathcal{M}_1(\mathcal{M}_0=0)=1$. In addition, it is easy to prove that $\forall \mathcal{M}_0$, $\partial (\mathcal{M}_0+\mathcal{M}_1)/\partial \mathcal{M}_0 > 0$,  so that $\mathcal{M}_0+\mathcal{M}_1 > 1$ is always verified.

Then $r=\rho_1/\rho_0$ follows from $r = 1+ \mathcal{M}_0/\mathcal{M}_1$.

For small impact velocities, namely $\mathcal{M}_0 \ll 1$, we find
\begin{eqnarray}
  r                             &=& 1 + \mathcal{M}_0 +  \mathcal{O}\left(\mathcal{M}_0^2\right) , \label{eq:weak} \\
  \mathcal{M}_1                 &=& 1 + \frac{\hat{\gamma} - 3}{4} \mathcal{M}_0 +  \mathcal{O}\left(\mathcal{M}_0^2\right), \nonumber \\
  \mathcal{M}_0 + \mathcal{M}_1 &=& 1 + \frac{\hat{\gamma} +1}{4} \mathcal{M}_0 +  \mathcal{O}\left(\mathcal{M}_0^2\right). \nonumber
\end{eqnarray}
In the opposite limit $\mathcal{M}_0 \gg 1$,
\begin{eqnarray}
  r                              & \rightarrow &  \frac{\hat{\gamma}+1}{\hat{\gamma}-1}~~\mathrm{(strong~shock~limit)}, \\
  \mathcal{M}_1                  & \rightarrow &   \frac{\hat{\gamma} - 1}{2} \mathcal{M}_0  ,  \nonumber  \\
  \mathcal{M}_0 + \mathcal{M}_1  & \rightarrow &   \frac{\hat{\gamma} + 1}{2} \mathcal{M}_0  . \nonumber
\end{eqnarray}
%


\end{document}